\documentclass{iopart}

	\usepackage{iopams}
	\usepackage{amsthm}
	\usepackage{setstack}
	\usepackage{graphicx}
	\usepackage{subfigure}

\newtheorem{mytheo}{Theorem}

\begin{document}

\title[Corrigendum to ``Thermodynamical instabilities in General Relativity"]{Corrigendum to ``Thermodynamical instabilities of perfect fluid spheres in General Relativity''}
\author{Zacharias Roupas$^{1}$}
\address{$^1$ Institute of Nuclear and Particle Physics, N.C.S.R. Demokritos, GR-15310 Athens, Greece} 
\ead{roupas@inp.demokritos.gr}

\begin{abstract}
	In \cite{Roupas:2013nt}, the thermal equilibrium of static, spherically symmetric perfect fluids in General
Relativity was studied. I would like to elaborate three points relevant to the results of [1]. The
first point is only a clarification, summarized in theorem 1 below, of results that appear in [1].
The following two points correct the error in [1], stating that the condition for thermodynamic
stability, found in [1], is referring to the microcanonical ensemble, while it was referring to
the canonical one. In theorems 2 and 3, specific cases for which equivalence of dynamical and
thermodynamic stability holds are specified.
\end{abstract}

\maketitle

\begin{enumerate}
\item The Tolman-Oppenheimer-Volkoff (TOV) equation (that is the equation one is led to from Einstein's equations and expresses the hydrostatic equilibrium) was derived in \cite{Roupas:2013nt} from extremization of the total entropy, for fixed total mass and number of particles. There was used the Lagrange multipliers method as in equation (28) of Ref. 
\cite{Roupas:2013nt}, namely:
\begin{equation}
	\delta S - \beta_0 \delta M c^2 + \alpha \delta N = 0
\end{equation}
where $\beta_0$, $\alpha$ were some as yet undetermined Lagrange multipliers. The first multiplier was found to be $\beta_0 = 1/T_0$ as in equation (33) of Ref. \cite{Roupas:2013nt}, where $T_0$ is the constant Tolman temperature, i.e. the surface temperature as measured by a distant observer. The second Lagrange multiplier was found to be 
\begin{equation}\label{eq:alpha}
	\alpha = \frac{\mu(r)}{T(r)} = const.
\end{equation}
where $\mu(r)$ is the chemical potential. 
Using equation (\ref{eq:alpha}) and standard thermodynamic relations, as in Ref. \cite{PhysRevD.85.027503}, it was shown that:
\begin{equation}\label{eq:dT}
	\frac{1}{T}\frac{dT}{dr} = \frac{1}{p+\rho c^2}\frac{dp}{dr}
\end{equation}
Using Einstein's equations it is straightforward to see that equation (\ref{eq:dT}) is the differentiated form of Tolman's relation:
\begin{equation}\label{eq:Tol_T}
	T(r)\sqrt{g_{tt}}=T_0 \equiv const.
\end{equation}
Indeed, for $g_{tt} = e^\nu$, equation (\ref{eq:Tol_T}) gives $T'/T = -\nu'/2$, which by use of equation (A.7) of Ref. \cite{Roupas:2013nt} that is derived from Einstein's equations, leads directly to (\ref{eq:dT}).
Therefore equation (\ref{eq:alpha}) gives:
\begin{equation}\label{eq:Tol_mu}
	\mu(r)\sqrt{g_{tt}} = \mu_0 \equiv const.
\end{equation}

Thus, in Ref. \cite{Roupas:2013nt} is shown that (see also \cite{Klein:1949}): 
\begin{mytheo}\label{th:equil}
For static spherically symmetric perfect fluids in General Relativity, thermal equilibrium requires the TOV equation to hold along with Tolman's relation (\ref{eq:Tol_T}) and equation (\ref{eq:Tol_mu}).
\end{mytheo} 
This was later generalized in \cite{Green:2013ica} for the stationary case. In addition, this calculation means that, \textit{at least in the case under study, thermal equilibrium implies dynamical equilibrium, while the inverse is not necessarily true}, since the Einstein's equations do not imply equations (\ref{eq:Tol_T}) and (\ref{eq:Tol_mu}).

\item Regarding the relation between thermodynamic and dynamical \textit{stability}, it was claimed in Ref. \cite{Roupas:2013nt} that they are equivalent for a perfect fluid under radial perturbations, provided relation (50) of Ref. \cite{Roupas:2013nt} holds, namely:
\begin{equation}\label{eq:Canonical}
	\frac{dp}{dr}\delta\rho = \delta p\frac{d\rho}{dr}.
\end{equation}
It must be stressed, however, that \textit{this is the condition for the dynamical stability to be equivalent with thermodynamic stability only in the canonical ensemble}, i.e. in the presence of a heat bath, and not in the microcanonical ensemble as was wrongly stated in \cite{Roupas:2013nt}. The reason is that in deducing condition (\ref{eq:Canonical}), the following equation (equation (44) of Ref. \cite{Roupas:2013nt}) was imposed:
\begin{equation}\label{eq:Can_T}
	\frac{\delta T}{T} = \frac{\delta p}{p+\rho c^2}
\end{equation}
However as was shown earlier, this equation follows directly from the condition
\begin{equation}\label{eq:Can_Tol}
	\delta T_0 = 0.
\end{equation}
Note also that the condition (\ref{eq:Canonical}) refers to constant number of particles. This constraint of perturbations that preserve the total number of particles is imposed in the second variation of entropy (46) in Ref. \cite{Roupas:2013nt}. 

In addition, I remark that the whole formulation includes also the case of zero chemical potential with no additional constraints, i.e. in any ensemble.

It is evident that equivalence between thermodynamic and dynamical stability depends not only on physical constraints of the system, such as the presence of a heat bath (canonical case) or the system being isolated (microcanonical case) but also on the equation of state. Let me formulate one theorem, that specify cases which meet canonical equivalence and one theorem that specify a case which meets general equivalence.
\begin{mytheo}\label{th:stab}
For static, spherically symmetric perfect fluids and for radial perturbations, dynamical stability in the presence of a heat bath is equivalent with thermodynamic stability in the canonical ensemble, in the cases:
\begin{enumerate}
	\item Constant entropy per particle $(s/n) = const.$
	\item $p = p(\rho,T)$
\end{enumerate}
where $s$, $n$, $\rho$ are the proper entropy, particle and mass density respectively, while $T$, $p$ are the proper temperature and pressure respectively. 
\end{mytheo} 

Note that case (ii) includes also the simpler, though popular, case $p = p(\rho)$. 

Let us prove case (i). We have for $p = p(\rho,n)$ that:
\begin{equation}\label{eq:pentropy}
\begin{array}{l}
\displaystyle 	p' = \frac{\partial p}{\partial\rho}\rho' + \frac{\partial p}{\partial n} n' 
\\[2ex]
\displaystyle	\delta  p = \frac{\partial p}{\partial\rho} \delta \rho + \frac{\partial p}{\partial n} \delta n 
\end{array}.
\end{equation}
Using (\ref{eq:pentropy}), the condition (\ref{eq:Canonical}) for equivalence in the canonical ensemble becomes:
\begin{equation}\label{eq:Con_can_1}
	n'\delta \rho = \rho '\delta n,
\end{equation}
Now from Euler relation, along with the condition (\ref{eq:alpha}) for thermal equilibrium we get:
\begin{equation}\label{eq:nprime}
		T s = p+\rho c^2 - \mu n \Rightarrow \frac{d}{dr}\frac{p+\rho c^2}{Tn} = 0 
\end{equation}
and substituting condition (\ref{eq:dT}) for thermal equilibrium  into (\ref{eq:nprime}) we finally get:
\begin{equation}\label{eq:rhoprime}
	n\rho' c^2 = (p+\rho c^2) n'
\end{equation} 
If the perturbation is made under conditions of constant Tolman temperature and number of particles, then equation (\ref{eq:Can_T}) holds and (\ref{eq:alpha}) gives:
\begin{equation}
	\delta\left( \frac{\mu}{T} \right)= 0 .
\end{equation}
Therefore, following similar steps we get also:
\begin{equation}\label{eq:rhodelta}
	n(\delta \rho) c^2 = (p+\rho c^2) (\delta n)
\end{equation} 
Dividing equations (\ref{eq:rhoprime}) and (\ref{eq:rhodelta}) by parts we are led to equation (\ref{eq:Con_can_1}). q.e.d.

	Let us prove case (ii). For $p = p(\rho,T)$, using the condition (\ref{eq:dT}) for thermal equilibrium, the pressure derivative may be written:
\begin{equation}\label{eq:pprime}
	p' = \frac{\partial p}{\partial\rho}\rho' + \frac{\partial p}{\partial T} T'
		= \left( 1-\frac{\partial p}{\partial T}\frac{T}{p+\rho c^2} \right)^{-1}\frac{\partial p}{\partial\rho}\rho'
\end{equation}
If the perturbation is made under conditions of constant Tolman temperature (note that the condition of constant number of particles is not needed), then equation (\ref{eq:Can_T}) holds and following similar steps the pressure variation may be written as:
\begin{equation}\label{eq:pdelta}
	\delta p = \left( 1-\frac{\partial p}{\partial T}\frac{T}{p+\rho c^2} \right)^{-1}\frac{\partial p}{\partial\rho}\delta \rho
\end{equation} 
Dividing equations (\ref{eq:pprime}) and (\ref{eq:pdelta}) by parts we are led to condition (\ref{eq:Canonical}). q.e.d.

\begin{mytheo}\label{th:zero}
For static, spherically symmetric perfect fluids with zero chemical potential $\mu = 0$ and for radial perturbations, dynamical stability is equivalent with thermodynamic stability.
\end{mytheo}

Indeed, for these systems the first law gives: $T\delta s = \delta\rho c^2$, while the Euler relation gives:
\[
	Ts = p + \rho c^2 \Rightarrow \delta p = s \delta T\Rightarrow \delta p = \frac{p+\rho c^2}{T}\delta T\Rightarrow \delta T_0 = 0
\]
Therefore, in this case the condition for thermodynamic stability is (\ref{eq:Canonical}). Since such systems do additionally satisfy condition (ii) their dynamical stability is equivalent with thermodynamic stability. q.e.d.

\item Finally, regarding the Newtonian limit, it shall be clear that the `transformation of ensembles' stated at section 5 of Ref. \cite{Roupas:2013nt} does not take place, but it was my misunderstanding of the fact that in deriving equation (58) of Ref. \cite{Roupas:2013nt}, the canonical condition $\delta T_0 = 0$ in the form of equation (\ref{eq:Can_T}) was assumed. Thus, in Ref. \cite{Roupas:2013nt}, it was proved that \textit{for static spherically symmetric perfect fluids with a linear equation of state, canonical thermodynamic stability in General Relativity gives canonical thermodynamic stability in the Newtonian limit}, as expected. Thermodynamic instabilities in General Relativity for the classical relativistic ideal gas are thoroughly discussed in \cite{Roupas:2014sda}, where the role of the Newtonian limit is very clearly outlined.

\end{enumerate}

\section*{Acknowledgements}

\indent I would like to thank Stephen Green, Joshua Schiffrin and Robert Wald for bringing into my attention the constant entropy per particle case.

\section*{References}

\bibliography{roupas_corrigendum_instability}
\bibliographystyle{h-physrev}

\end{document}